\documentclass[sigconf]{acmart}

\makeatletter
\def\@ACM@checkaffil{
    \if@ACM@instpresent\else
    \ClassWarningNoLine{\@classname}{No institution present for an affiliation}%
    \fi
    \if@ACM@citypresent\else
    \ClassWarningNoLine{\@classname}{No city present for an affiliation}%
    \fi
    \if@ACM@countrypresent\else
        \ClassWarningNoLine{\@classname}{No country present for an affiliation}%
    \fi
}
\makeatother

\usepackage{graphicx}

\usepackage{amsthm}
\usepackage[linesnumbered,ruled]{algorithm2e}
\usepackage{comment}
\usepackage{subcaption}
\usepackage{enumerate}
\usepackage{mathtools}
\usepackage{comment}
\usepackage{paralist} 
\usepackage{multicol}
\usepackage{multirow}
\usepackage{makecell}
\usepackage{verbatim}
\usepackage{color}
\usepackage{float}

\definecolor{darkgrn}{rgb}{0, 0.75, 0}

\usepackage{xargs}                      

\newcommand{\covid}{COVID-19}
\newcommand{\covasim}{\texttt{Covasim}}
\newcommand{\preempt}{\texttt{PREEMPT}}

\usepackage[capitalize]{cleveref}

\setcopyright{acmcopyright}
\copyrightyear{2021}
\acmYear{2021}
\acmDOI{xx.xxxx/xxxxxxxxx.xxxxxxx}

\acmConference[epiDAMIK 2021]{4th epiDAMIK ACM SIGKDD International Workshop on Epidemiology meets Data Mining and Knowledge Discovery}{Aug 15, 2021}{Virtual}
\acmBooktitle{epiDAMIK 2021: 4th epiDAMIK ACM SIGKDD International Workshop on Epidemiology meets Data Mining and Knowledge Discovery}
\acmPrice{15.00}
\acmISBN{978-1-xxxx-XXXX-X}

\begin{document}

\title{An Integrated Epidemic Simulation Workflow for Submodular Intervention Strategies}

\author{Reet Barik}
\affiliation{Washington State University, Pullman, WA, USA}
\email{reet.barik@wsu.edu}

\author{Marco Minutoli}
\affiliation{Pacific Northwest National Laboratory, Richland, WA, USA}
\email{marco.minutoli@pnnl.gov}

\author{Mahantesh Halappanavar}
\affiliation{Pacific Northwest National Laboratory, Richland, WA, USA}
\affiliation{Washington State University, Pullman, WA, USA}
\email{hala@pnnl.gov}

\author{Ananth Kalyanaraman}
\affiliation{Washington State University, Pullman, WA, USA}
\affiliation{Pacific Northwest National Laboratory, Richland, WA, USA}
\email{ananth@wsu.edu}

\begin{abstract}

Owing to the ongoing \covid{} pandemic and other recent global epidemics, epidemic simulation frameworks are gaining rapid significance. In this work, we present a workflow that will allow researchers to simulate the spread of an infectious disease under different intervention schemes. Our workflow is built using the \covasim{} simulator for \covid{} alongside a network-based \preempt{} tool for vaccination. The \covasim{} simulator is a stochastic agent-based simulator with the capacity to test the efficacy of different intervention schemes. \preempt{} is a graph-theoretic approach that models epidemic intervention on a network using submodular optimization. By integrating the \preempt{} tool with the \covasim{} simulator, users will be able to test network diffusion based interventions for vaccination. The paper presents a description of this integrated workflow alongside preliminary results of our empirical evaluation for \covid. 


\end{abstract}

\keywords{epidemic simulation, \covid, vaccination, intervention, influence maximization}

\maketitle
\pagestyle{plain}
\section{Introduction}
\label{sec:introduction}

Largely spurred by the ongoing \covid{} pandemic, the past year has seen a dramatic growth in the demand for effective epidemic simulators (see Section~\ref{sec:RelatedWork} for a selective subset of related works). 
Most of these simulators implement agent-based compartmentalized models for epidemic simulations and are effective at simulating disease progression over time. 
While many of these simulators also support the ability to specify various intervention schemes (vaccinations, social distancing, closures, etc.), these intervention schemes are typically specified at the coarse level of the network---e.g., as probability variables to encode contact probability or transmission probability. 

However, recent algorithmic developments in network science have introduced an alternative way to specify intervention. For instance, treating the disease spread on an epidemic (contact) network as a diffusion process and vaccination as a node removal process, one can study the impact of varying vaccination schemes at finer granularities of the network, and based on the network characteristics of the nodes to remove. 
Existing epidemic simulation platforms lack the  ability to plug-and-play such network-based intervention mechanisms. 

{\bf Contribution:} 
In this paper, we present an integrated simulation workflow to carry out epidemic simulations with network-based interventions. In particular, we focus on vaccination schemes for interventions. For simulator, our workflow uses the \covasim{} agent-based simulator \cite{kerr2020covasim} for \covid. For vaccination-based intervention, our workflow uses the \preempt{} method that uses submodular optimization for identifying nodes to vaccinate on a network \cite{DBLP:conf/sc/MinutoliSHTKV20}.
For comparative purposes, we also implemented simpler schemes such as random and degree-based seed selection.

\begin{figure*}[ht!]
    \centering
    \includegraphics[width=0.75\textwidth]{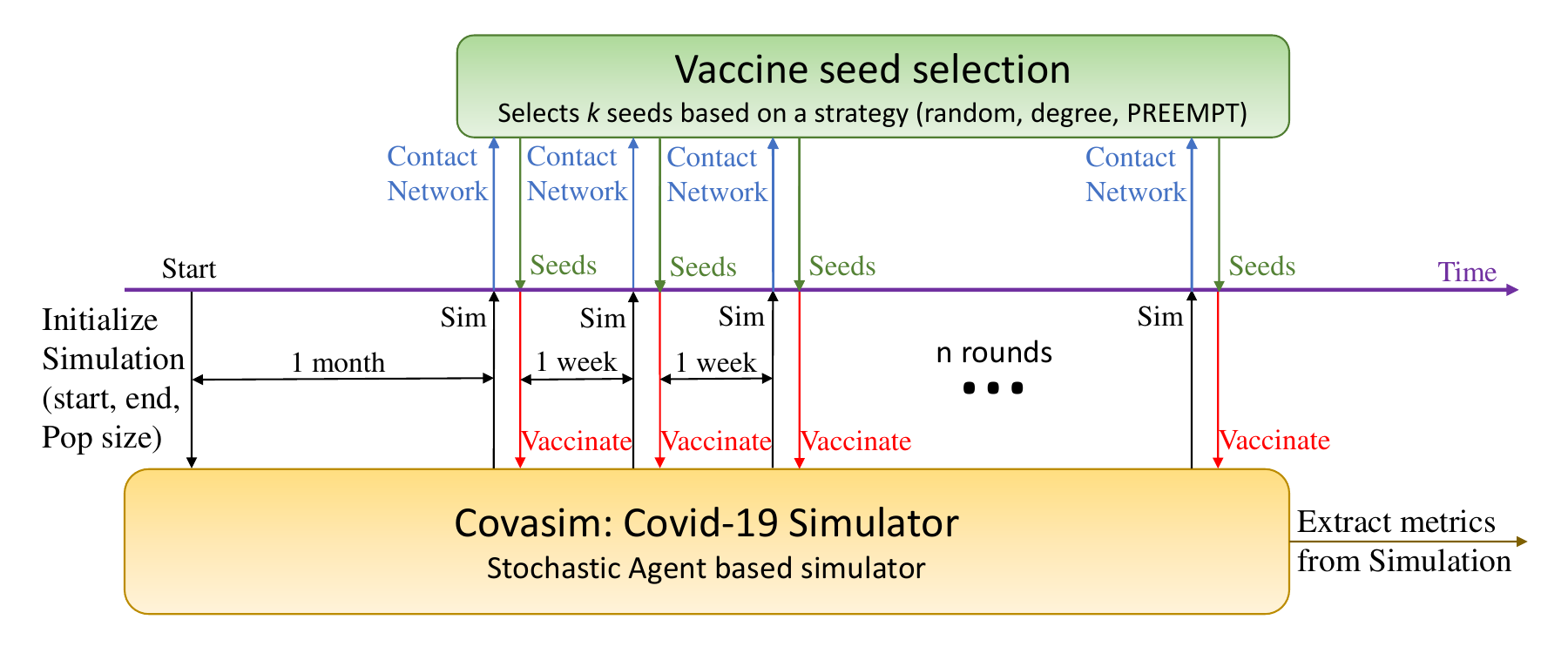}
    \caption{
    \small
    \textit{Our integrated workflow for submodular epidemic intervention:} The simulation is allowed to run unhindered for a month followed by regular vaccination rounds of certain batch sizes every week. Nodes to vaccinate are specified by a seed selection strategy, which could internally implement various strategies to identify those seeds. 
    }
    \label{fig:workflow}
    \vspace{-0.2cm}
\end{figure*}

The simulation workflow functions as follows, and is schematically illustrated in Figure~\ref{fig:workflow}. 
The simulations typically cover a span of time of the disease spread (from days to months). During this time, the simulator uses an in-built diffusion model to simulate the spread of the disease over the population. Our workflow will allow users to intervene at various stages of the simulation, by specifying which subset of nodes (or individuals on the contact network) to vaccinate at that stage. We refer to these regular intervention stages as ``vaccination rounds''. 
The simulator uses this intervention to update the underlying contact network, and then proceeds with the simulation. The simulator allows various network measurements that vary temporally (e.g., number of infected nodes, number of hospitalizations, number of deaths). These measurements are tracked so that we can thoroughly study the efficacy of different vaccination schemes. 

In Section~\ref{sec:RelatedWork}, we cover relevant works on epidemic simulators.
In Section~\ref{sec:infmax} we present the algorithmic foundations for vaccination-based intervention. 
In Section~\ref{sec:ExpFramework} we describe the key components of our integrated workflow implemented for this effort, including the necessary background on the \covasim{} simulator and \preempt{} tool. 
Section~\ref{sec:ExpEval} presents the results of our experimental evaluation and comparisons.  

\section{Related Works on Epidemic Simulators}
\label{sec:RelatedWork}
Some epidemic simulators like CovidSim \cite{ferguson2020covid} use a spatial model that divides a specific geographical location into cells and simulates the spread of the disease based on the evolving properties of each cell. Compartmental models on the other hand like the classical SIR model \cite{kermack1927contribution} or its more sophisticated counterparts like SEIR, SEIRS, SIRS, SEI, SEIS, SI, and SIS \cite{hethcote2000mathematics} provide the option to model the disease spread  through ordinary differential equations or in a stochastic manner. CovidSIM \cite{schneider2020covid} is one such simulator which models the spread of \covid{} in a deterministic way using an extended SEIR model. A network based stochastic epidemic simulator was introduced in \cite{kuzdeuov2020network} where a country is represented as a graph with individual nodes representing an administrative unit of the country, such as a city or region. SC-COSMO \cite{kunst2020stanford} and COVID-19 Simulator~\cite{covid19-sim} fall under the category of simulators which support the specification of interventions at a coarser level. A more fine grained view is presented by agent-based simulators which have been used to simulate outbreaks in the past. For instance, an agent-based simulator like~\cite{frias2011agent} that uses social interactions and individual mobility patterns was used to study the H1N1 outbreak in Mexico. 
More such agent-based stochastic simulators that were historically used to simulate the spread of diseases like Measles have been adapted to be used for \covid{} like FRED \cite{grefenstette2013fred}. 

\section{Influence Maximization-based Intervention}
\label{sec:infmax}

Influence maximization \cite{DBLP:conf/sigmod/TangSX15,DBLP:conf/kdd/KempeKT03} is a problem that has been well studied in social networks.
We are given a graph $G(V,E)$ of $n$ nodes and $m$ edges, and a diffusion process that dictates how the information is spread from node to node,
the classical problem \cite{DBLP:conf/kdd/KempeKT03} is one of identifying a set of $k$ nodes that is expected to maximize the influence spread based the diffusion process.

Targeted immunization on the other hand, is the problem to identify nodes to vaccinate in a given network, such that the expected disease spread is minimized. Under the Linear Threshold diffusion model, these two problem have been shown to be equivalent in \cite{cheng2020outbreak}. As a result, influence maximization based framework as an approximate solution for designing intervention (specifically vaccination) strategies is a feasible method for epidemic control.

The framework of networked epidemiology parts from the standard, but computationally more efficient, compartmental mass action models \cite{chen2020networked} by providing a framework that can capture the dynamics of an epidemic spreading over a population more precisely.  In this framework, interactions among the individuals of the population are modeled through a network $G=(V,E,w)$. Each edge of the network $e=(u,v) \in E$ represent the interaction between individuals $u,v \in V$ and the weight function $w(e)$ provides the probability of transmitting the disease during that interaction. Therefore, the networked epidemiology framework provides a platform to design and test intervention policies at the individual level by taking into account the role of individuals in the contact network from both a functional and topological perspective.  In this context, the computational task of defining an optimal intervention strategy can be formulated as the following optimization problem:

\begin{definition}[EpiControl]
Given a contact network $G=(V,E,w)$, a diffusion model $M$ over $G$, a set of initially infected vertices $B \subseteq V$, and a fixed budget $k$, the EpiControl problem is to find an intervention set $S \subseteq V$ of size $k$ such that the expected number of infections at the end of the diffusion process ($\sigma(B, S)$) is minimized. 
\end{definition}

Computing $\sigma(B,S)$ can be approximated by choosing a sampling effort $N$ and averaging over these $N$ observations. More precisely, we can obtain a set $\{G_1\ldots G_N\}$ of subgraphs of $G$ capturing $N$ realization of the diffusion process $M$ by edge deletion over $G$: each subgraph $G_i$ will retain only the edges that have enabled the transmission of the disease during the simulation of the diffusion process.

For each of the subgraph $G_i$, the number of infection at the end of the diffusion process can be obtained by computing the function:
\begin{equation}
\sigma_{G_i}(B,S) = \left|\bigcup_{b \in B} R(b, S)\right|
\end{equation}
where $R(b, S) \subseteq V $ provides the set of vertices of reachable on $G_i$ from $b \in B$ when $S$ is our intervention set.  In the rest of our presentation, we will make the assumption that the vertices in $S$ will have perfect immunity from the disease.  Therefore, the function $R(b,S)$ over the sample $G_i$ can be computing by deleting the vertices in $S$ from $G_i$ and the performing a breadth-first search from $b$.

The main objective of this work is to provide a framework that can be used to experiment with various seed selection strategies (identifying which nodes to vaccinate in an input contact network) and evaluate their effectiveness in reducing the spread of an infectious disease like \covid{}. For this purpose we use an agent-based simulator that models the spread of \covid{} called \covasim{} \cite{kerr2020covasim}. The setup itself is generic enough so that other simulators can be used in its place.
\section{Experimental Framework}
\label{sec:ExpFramework}

This section aims to describe the suggested framework by giving a brief overview of the components, starting with the simulator used: \covasim{}.

\subsection{\texttt{Covasim}}
\label{sec:covasim}

\covasim{} \cite{kerr2020covasim} is a stochastic agent-based simulator that is used to simulate the spread of the Covid-19 disease. The other functionality of \covasim{} that is relevant to this work is the ability to define and evaluate different intervention strategies, more specifically, vaccines. The underlying model used by the simulator to generate a synthetic population in the form of a network for a given geographical location is based on works like \cite{prem2017projecting} to reflect age-specific mixing patterns for a specific population. The network itself is a multi-layered one with different layers (household, school, work, community) describing how the agents (nodes) interact with each other (edges). Every pair of adjacent nodes $(u,v)$ in the network have one edge from $u$ to $v$ and another from $v$ to $u$. The weights on the edges are the probability of the source node infecting the destination node given that the source node itself is infected. The probability value itself is modeled as a product of terms representing the infectiousness of the disease, the transmissibility of the source, the susceptibility of the destination, and the frequency of contact between the two. 

One of the seed selection strategies that was incorporated into the framework is the one given by \cite{DBLP:conf/sc/MinutoliSHTKV20} based on the near equivalence of the \textit{EpiControl} and Influence Maximization problem. This strategy, henceforth referred to as \preempt{}, is described below. 

\subsection{\preempt{}}
\label{sec:preempt}

\citet{DBLP:conf/sc/MinutoliSHTKV20} propose \preempt{} an intervention strategy that emerges from a new formulation of the EpiControl problem enabling to transfer some of the algorithmic techniques that have been developed for the maximization of submodular function under cardinality constraints.  They note that the objective function of the EpiControl problem can be reformulated in terms of the number of individuals spared from the disease through the intervention.  More precisely, they define the function to compute the \emph{number of lives saved} from the disease as:
\begin{equation}
\label{eq:lives-saved}
\lambda_{G_i}(B,S) = \sigma_{G_i}(B, \emptyset) - \sigma_{G_i}(B, S).
\end{equation}
Intuitively, \cref{eq:lives-saved} defines the difference in the number of infection between taking no action and deploying the intervention $S$. \citet{DBLP:conf/sc/MinutoliSHTKV20} also show the following theorem.
\begin{theorem}
Given a graph $G_i=(V,E)$, a set of initially infected individuals $B \subseteq V$, and an intervention set $S$, the function $\lambda_{G_i}$ of \cref{eq:lives-saved} is a submodular function of $S$ if $G_i$ is a rooted tree.
\label{th:submodularity}
\end{theorem}
Therefore, \citet{DBLP:conf/sc/MinutoliSHTKV20} reformulated the EpiControl problem as a maximization problem as follows:
\begin{definition}
Given a contact network $G=(V,E,w)$, a diffusion model $M$ over $G$, a set of initially infected vertices $B \subseteq V$, and a fixed budget $k$, the EpiControl problem is to find an intervention set $S \subseteq V$ of size $k$ such that the expected number of lives saved at the end of the diffusion process ($\lambda(B, S)$) is maximized. 
\end{definition}

The result of \cref{th:submodularity} and the formulation of the EpiControl problem suggest that the approximation algorithms that have been developed for the Influence Maximization\cite{DBLP:conf/sigmod/TangSX15,DBLP:conf/kdd/KempeKT03} problem can be adapted to solve the EpiControl problem when the samples $G_i$ are rooted trees. The result of \cref{th:submodularity} does not hold for the general case, but it includes the SIR model\cite{kermack1927contribution} and the class of diseases that evolve over trees (e.g., sexually transmitted diseases\cite{bearman2004chains}).  However, \citet{DBLP:conf/sc/MinutoliSHTKV20} note that when the structure of the samples $G_i$ is expected to be sparse and with few cycles the approximation algorithms for Influence Maximization lose their approximation guarantee, but they can serve as good heuristics.

\section{Experimental Evaluation}
\label{sec:ExpEval}

\subsection{Experimental Setup}
\label{sec:ExpSetup}

The following experimental setup was used to test our integrated workflow presented in Section~\ref{sec:ExpFramework} (and illustrated in Figure~\ref{fig:workflow}).

\begin{itemize}
    \item The duration of each run of the simulation is 170 days (over 5 months) -- starting on January 1st and ending on June 19th. 
    \item The simulation is allowed to run for the first 30 days unhindered. The first round of vaccination is on the 31st day. 
    Subsequent vaccination rounds occur once every week. We define a \textit{batch size} of a specific round to be the number of vaccines given at that round. 
    \item The contact network is extracted from the simulation framework and fed as input to the seed selection module to identify nodes to vaccinate. 
    \item The different types of vaccination strategies are: 
    \begin{itemize}
        \item \underline{Random}: The $k$ nodes to be vaccinated at a given intervention round are chosen at random. 
        \item \underline{Degree}: The next batch of top-$k$ high degree nodes are chosen to be vaccinated at each round.
        \item \underline{\preempt}: The top $k$ nodes identified by \preempt{} \cite{DBLP:conf/sc/MinutoliSHTKV20} based on the current state of the network, are vaccinated at a given  round. 
    \end{itemize}
\end{itemize}

\subsection{Experimental Results}
\label{sec:ExpResults}

Given the above setup we conduct three kinds of studies \footnote{An important thing to note while looking at the experimental results: The disease spread being simulated happens in a setting where the agents involved are unaware of the disease and there is no effect on their behavior (social distancing, reduced mobility etc.). This is why the relative performance of different strategies should be looked at rather than the inflated and unrealistic raw numbers.} to understand how different vaccination strategies (in terms of the number and type of vaccines given out) behave and use the no-vaccination strategy as the baseline for comparison. The statistics of the input contact network are as follows: 

\begin{itemize}
    \item Simulated location: India
    \item $|V| = 100,000$
    \item $|E| = 3,793,826$
    \item Degree distribution is shown in Figure~\ref{fig:degree}.    
    \vspace{-0.2cm}
    \begin{figure}[H]
    \centering
    \includegraphics[width=0.475\textwidth]{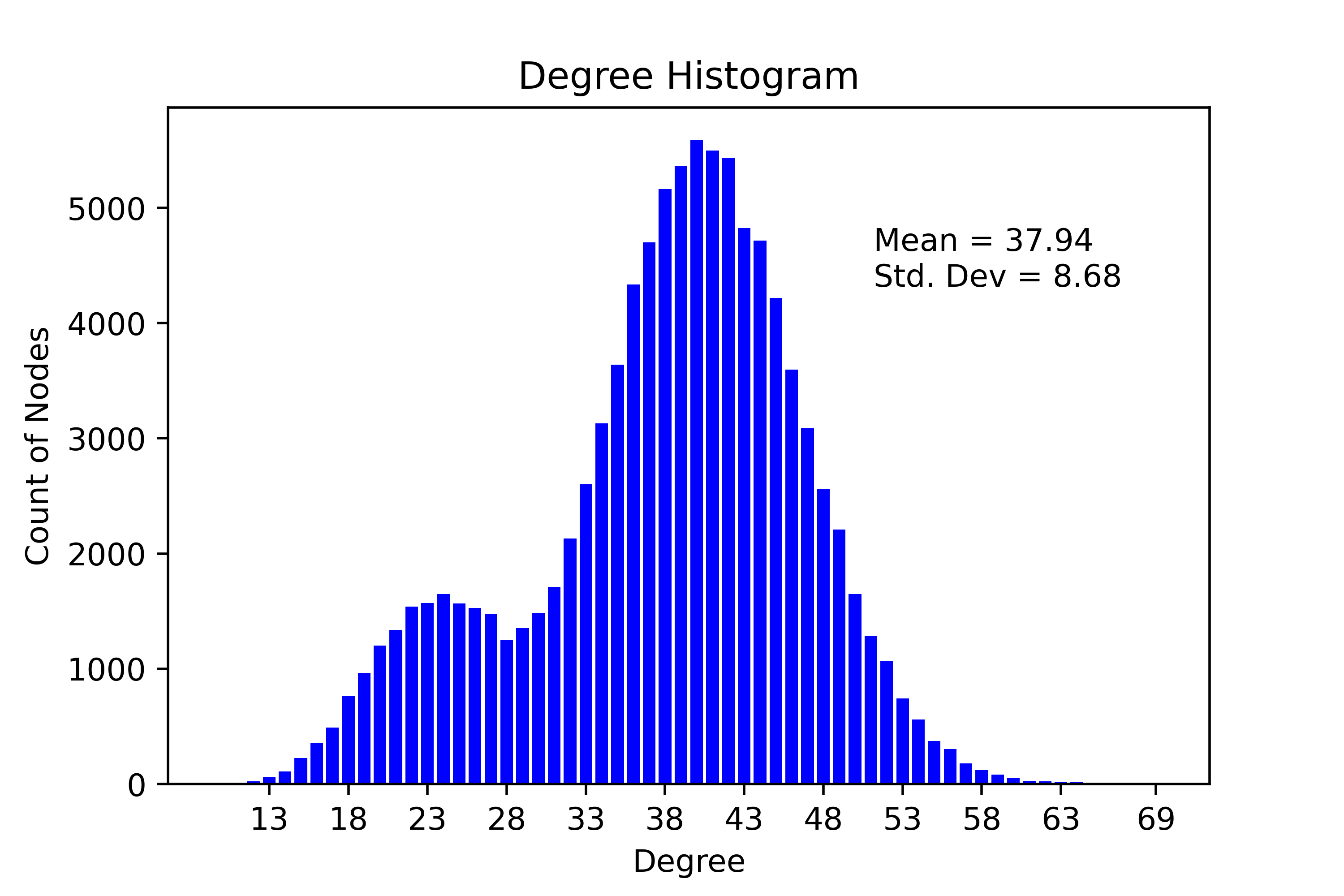}
    \vspace{-0.5cm}
    \caption{
    \small
    \textit{Degree Distribution of the Contact Network:} Since every adjacent node pairs have 2 edges between them, the figure below takes into account just the in-degree (which is same as the out-degree).}
    \label{fig:degree}
    \vspace{-0.2cm}
\end{figure}
\end{itemize}

The metrics we considered to evaluate different vaccination strategies are:
\begin{itemize}
    \item the number of \textit{cumulative infections} (i.e., since first day of simulation);
    \item The number of \textit{new infections per day}; and
    \item The number of \textit{cumulative deaths}.
\end{itemize}

In what follows, we use the term ``seeds'' to mean the nodes chosen for vaccination at any round.

\subsubsection{Effect of seed selection strategies}
\label{sec:singleround}

\begin{figure}[h!]
    \centering
    \includegraphics[width=0.475\textwidth]{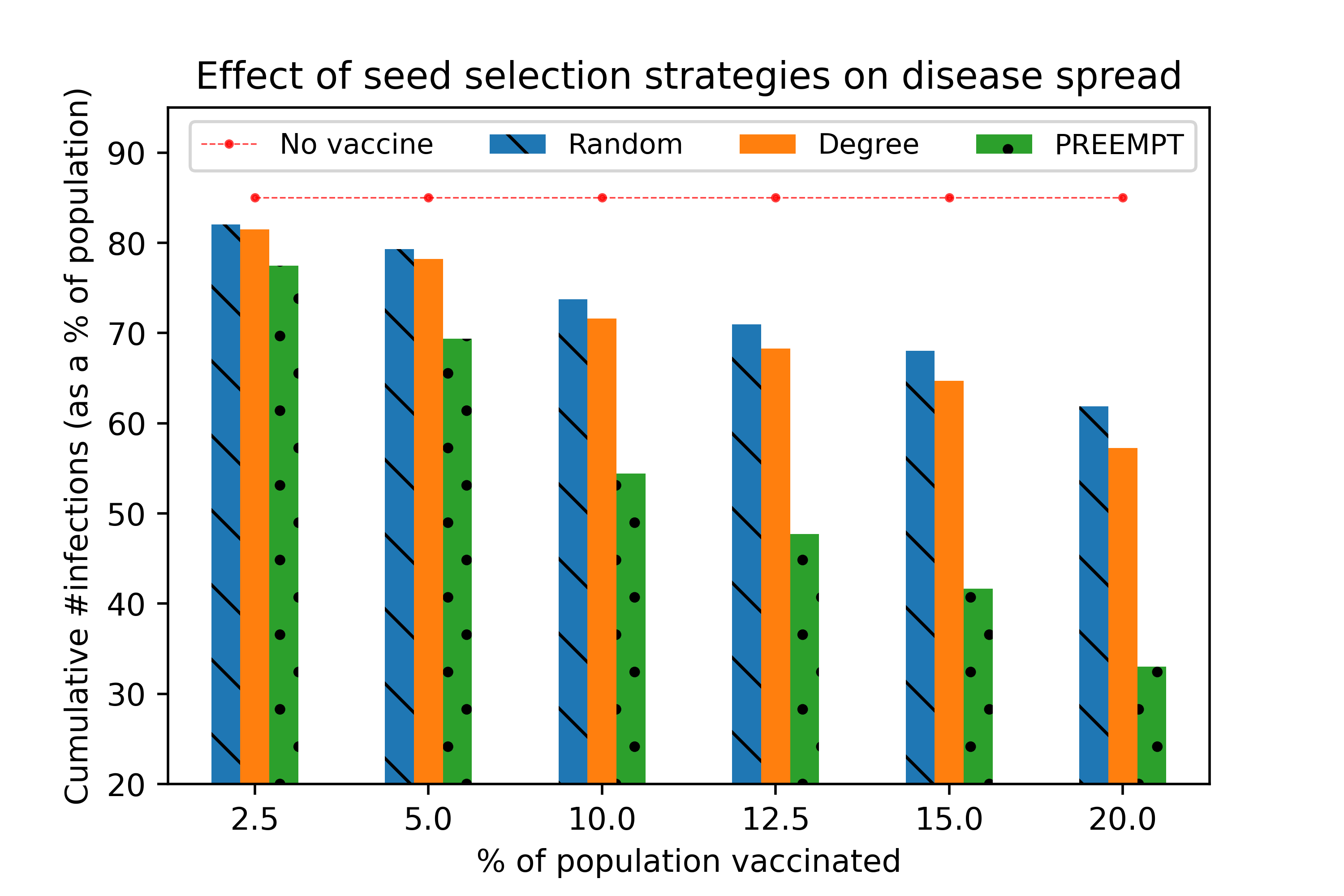}
    \caption{
    \small
    \textit{Effects of the seed selection strategy on disease spread:} The x-axis represents the \% of population vaccinated at a single round, on the 31st day of the simulation. The y-axis represents the cumulative \#infections after 5+ months of simulation as a \% of the population infected. }
    \label{fig:histogram}
    \vspace{-0.2cm}
\end{figure}

First, we evaluate the impact of the seed selection strategy on  the cumulative infections over a 5+ month period. For this, we vaccinate nodes prescribed by three strategies mentioned in Section ~\ref{sec:ExpSetup}, namely, Random, Degree, and \preempt{}. For each instance of the experiment, a fraction of the population is vaccinated based on these strategies at a \emph{single} round, on January 31st. The effects on the disease spread are observed in terms of cumulative infections over the course of the simulations. The results are shown in Figure ~\ref{fig:histogram}. 

Figure~\ref{fig:histogram} shows that vaccination strategies matter. 
More specifically, compared to the baseline of no-vaccination, all three strategies show reductions in the number of cumulative infections. 
However, the figure also helps show the varying efficacies among the different vaccination schemes.
In particular, \preempt{} significantly outperforms both random and degree-based seed selection strategies. For instance, when 20\% of the population is vaccinated, \preempt{} results in a 61\% decrease in the cumulative infections relative to the no-vaccine strategy; whereas the reductions achieved by the Random and Degree strategies are modest. 
In fact, the Degree based strategy shows only a marginal advantage over Random.
For this reason, henceforth, we will simply show results using Degree for comparison against \preempt.

In the above experiments, only a single round of vaccination was used, which may be unrealistic in practice. 
In what follows, we study the impact of temporally spacing out the vaccine delivery across multiple rounds. 
There are two configurations here to experiment---a) \textit{uniform}, where each round gets the same (fixed) number of vaccines (Section~\ref{sec:uniformbatches}); and b) \textit{non-uniform}, where the batch sizes can differ round to round (Section~\ref{sec:nonuniformbatches}). 


\subsubsection{Effect of vaccinating in batches of uniform size}
\label{sec:uniformbatches}

\begin{figure*}[tbh]
    \centering
    \begin{subfigure}[b]{.33\textwidth}
         \centering
         \includegraphics[width=\textwidth]{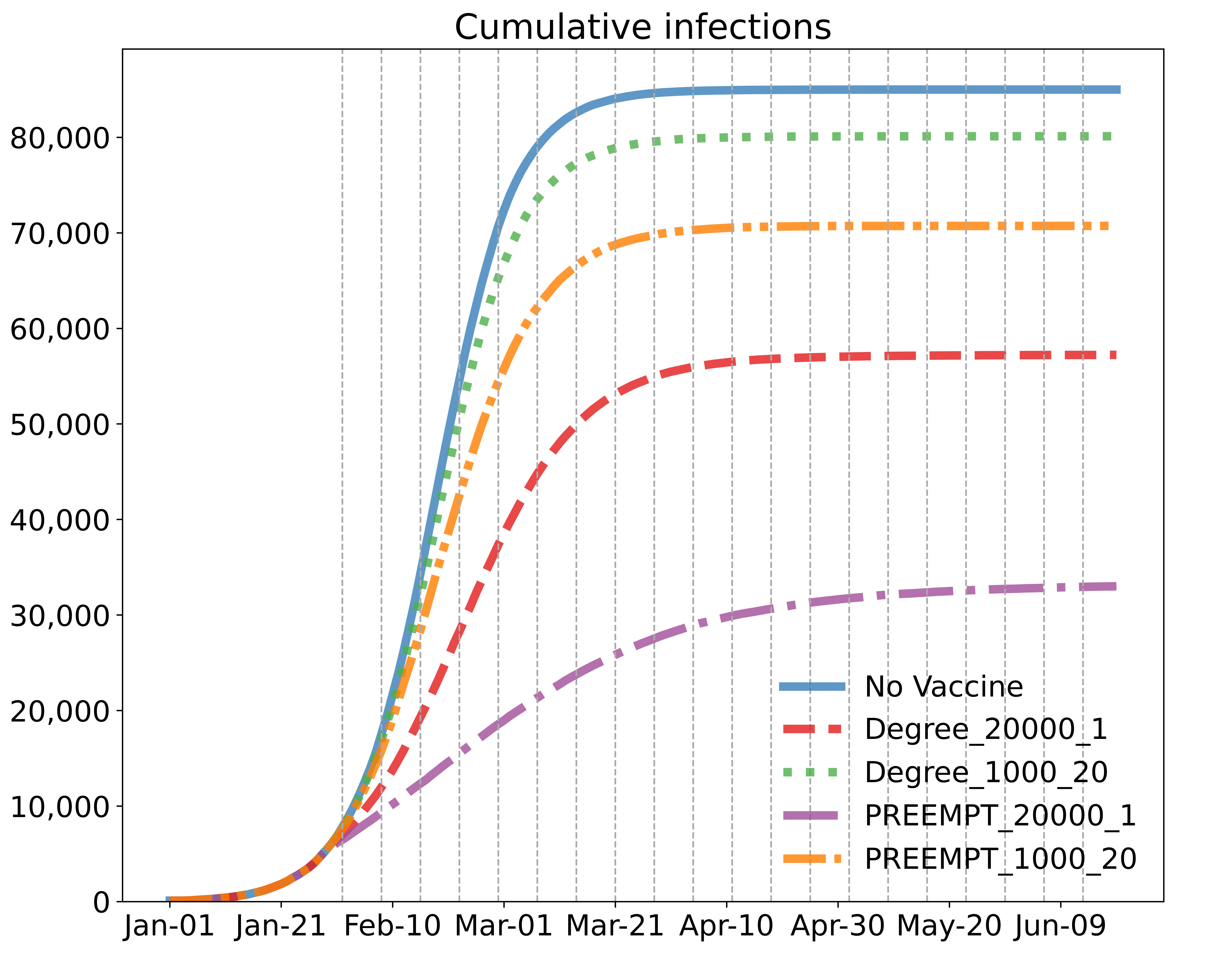}
         \caption{}
         \label{fig:4CI}
     \end{subfigure}
     \hfill
     \begin{subfigure}[b]{.33\textwidth}
         \centering
         \includegraphics[width=\textwidth]{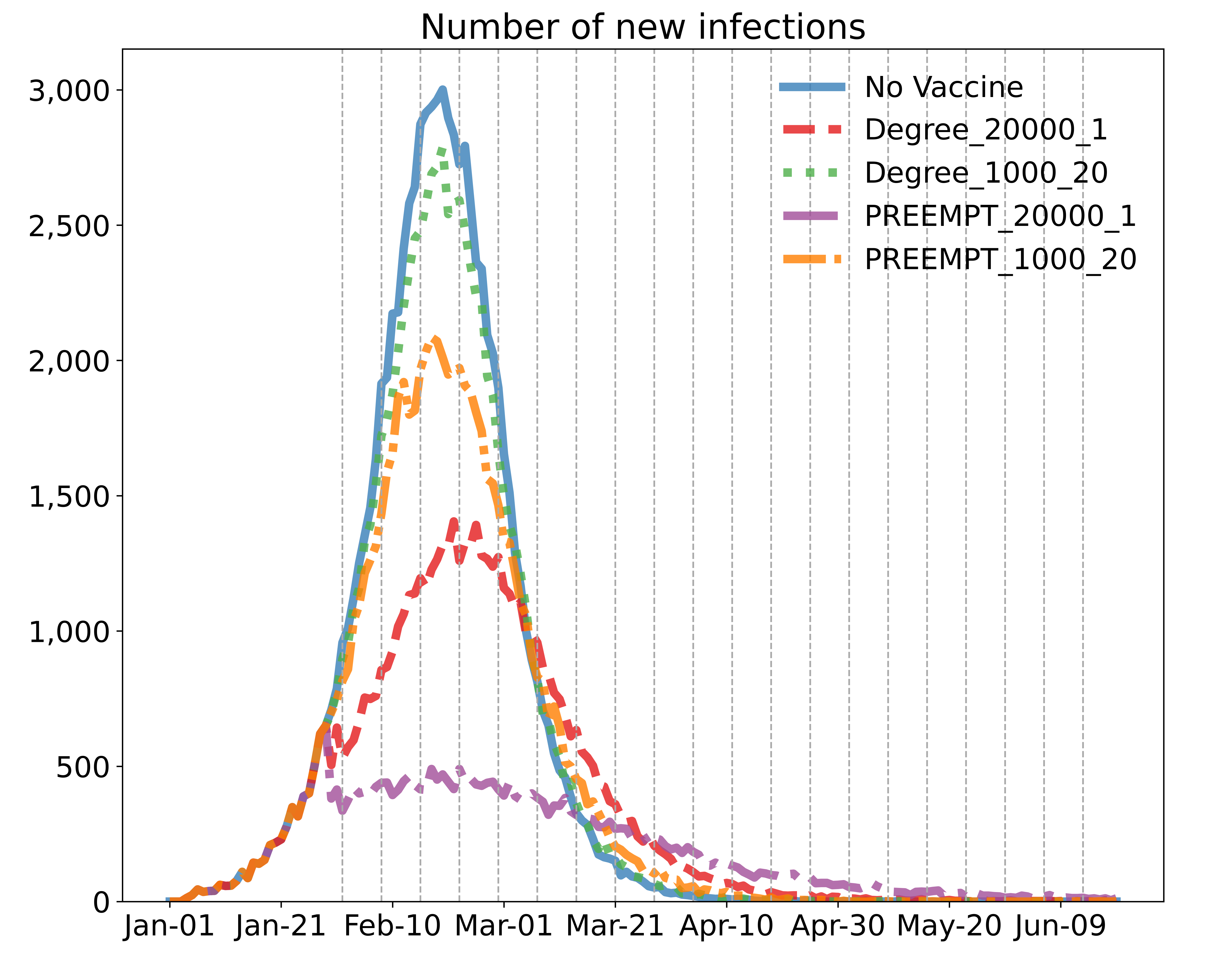}
         \caption{}
         \label{fig:4NI}
     \end{subfigure}
     \hfill
     \begin{subfigure}[b]{.33\textwidth}
         \centering
         \includegraphics[width=\textwidth]{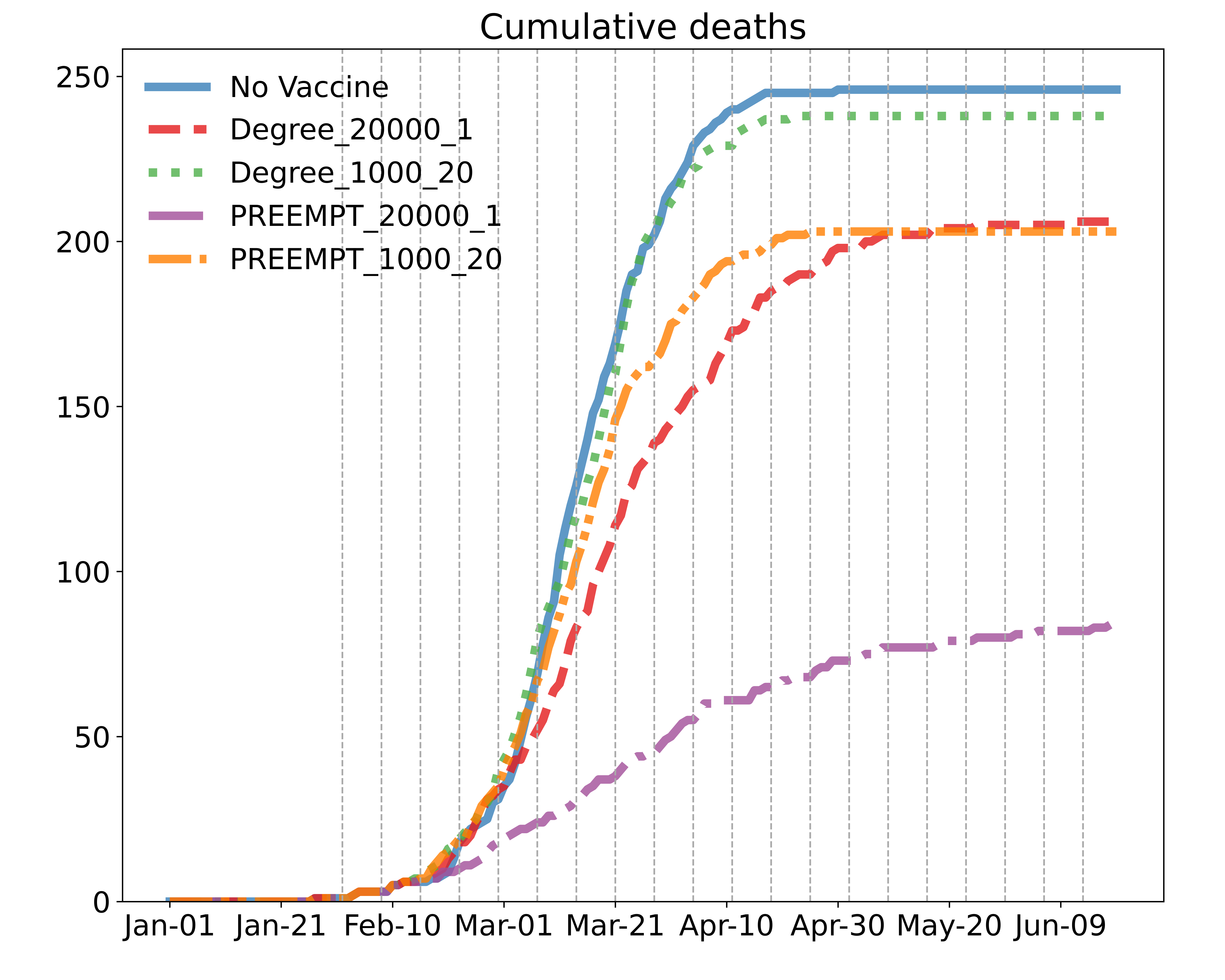}
         \caption{}
         \label{fig:4CD}
     \end{subfigure}
    \caption{
    \small
    \textit{Effects of temporally spacing out vaccination rounds on disease spread using a uniform batch size:} 
    For every curve labeled as `X\_Y\_Z', `X' stands for the seed selection strategy; `Y' stands for the (uniform) batch size used at each round; and `Z' stands for the number of rounds. 
    Every vertical dashed line represents a vaccination round.
    Also shown for comparative reference, is the `No vaccine' curve that corresponds to zero vaccines given out at each round. 
    We use 20,000 as the total number of vaccines.
    }
    \label{fig:uniformvaccination}
    \vspace{-0.2cm}
\end{figure*}

\begin{figure*}[h!]
     \centering
     \begin{subfigure}[b]{.33\textwidth}
         \centering
         \includegraphics[width=\textwidth]{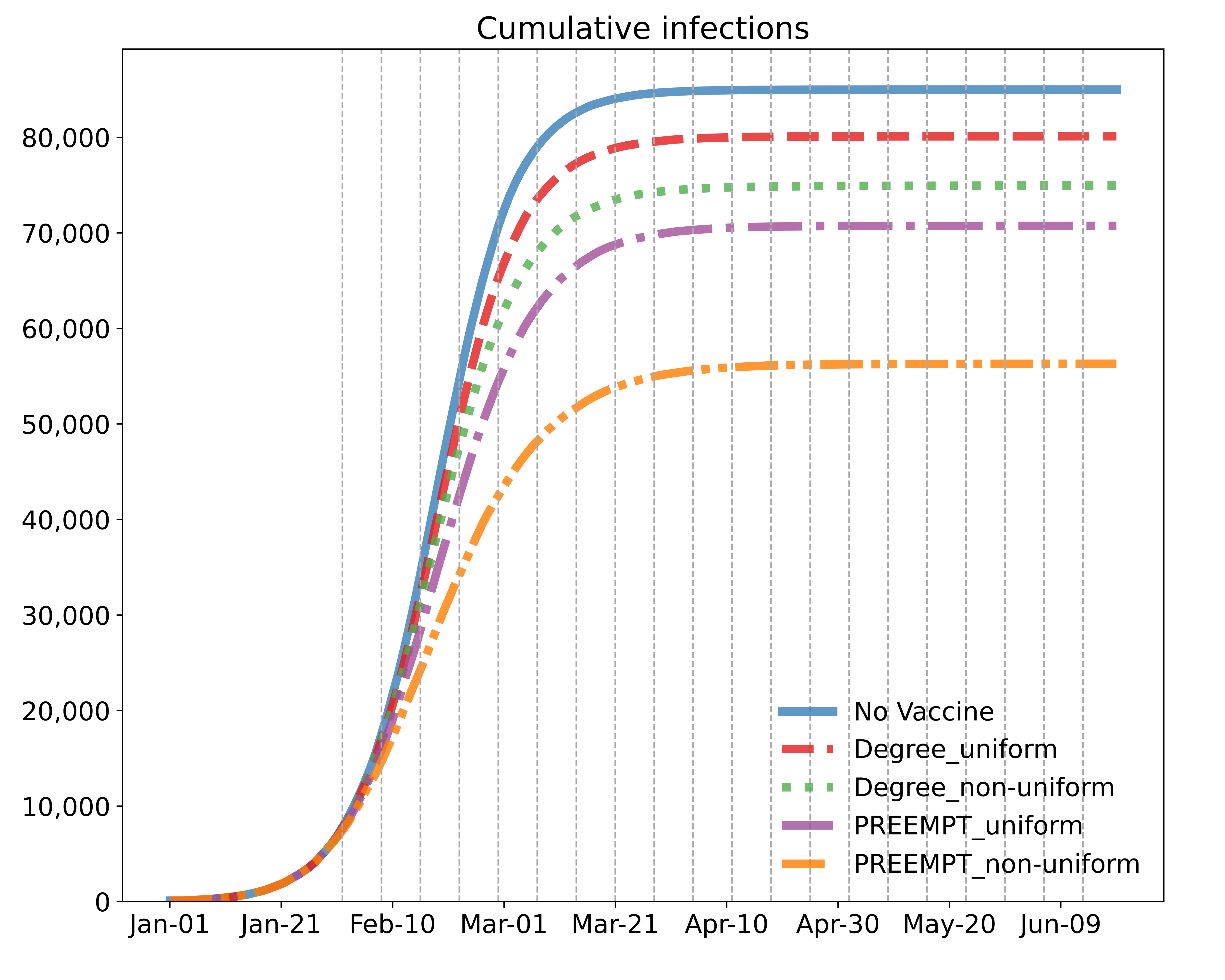}
         \caption{}
         \label{fig:5CI}
     \end{subfigure}
     \hfill
     \begin{subfigure}[b]{.33\textwidth}
         \centering
         \includegraphics[width=\textwidth]{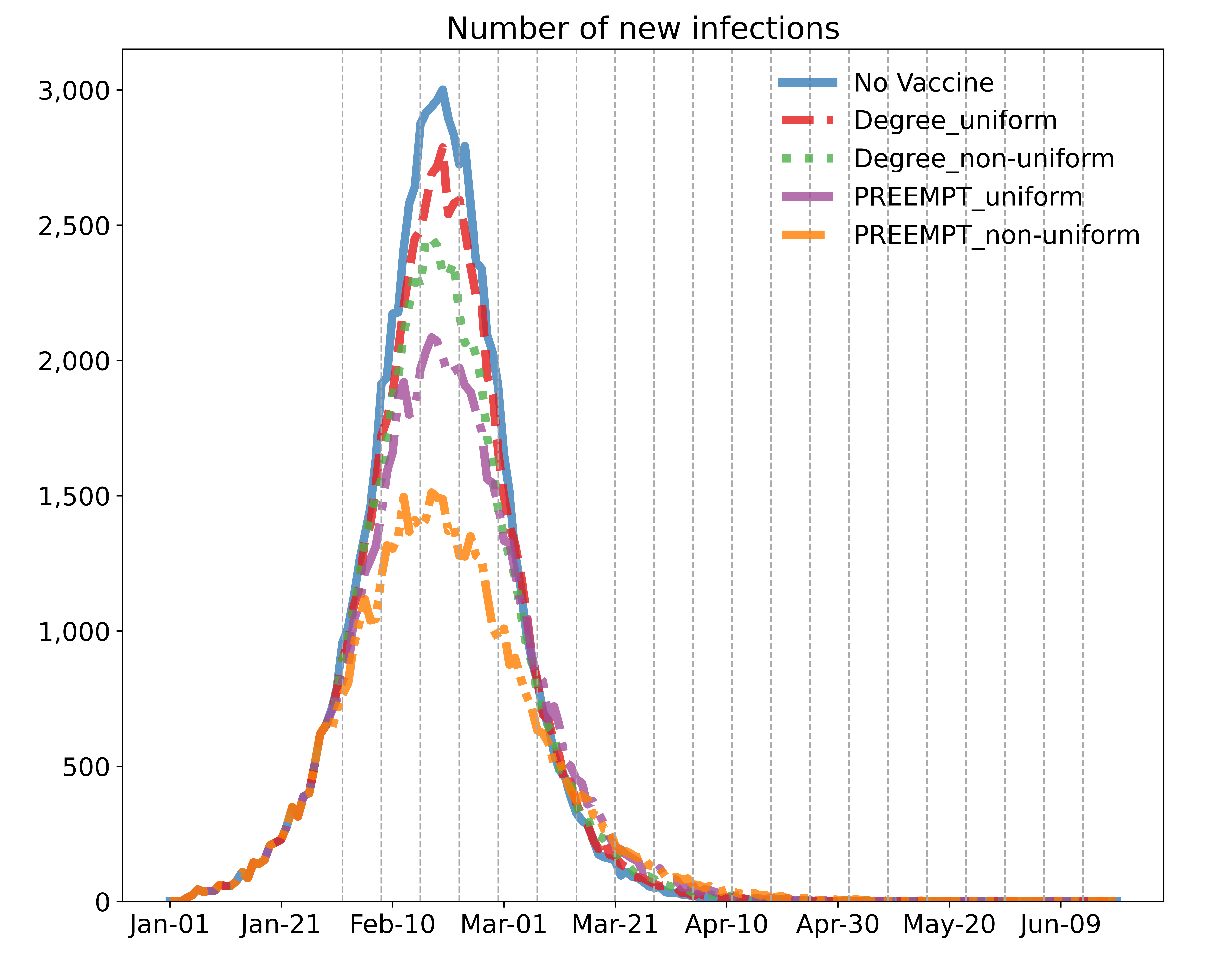}
         \caption{}
         \label{fig:5NI}
     \end{subfigure}
     \hfill
     \begin{subfigure}[b]{.33\textwidth}
         \centering
         \includegraphics[width=\textwidth]{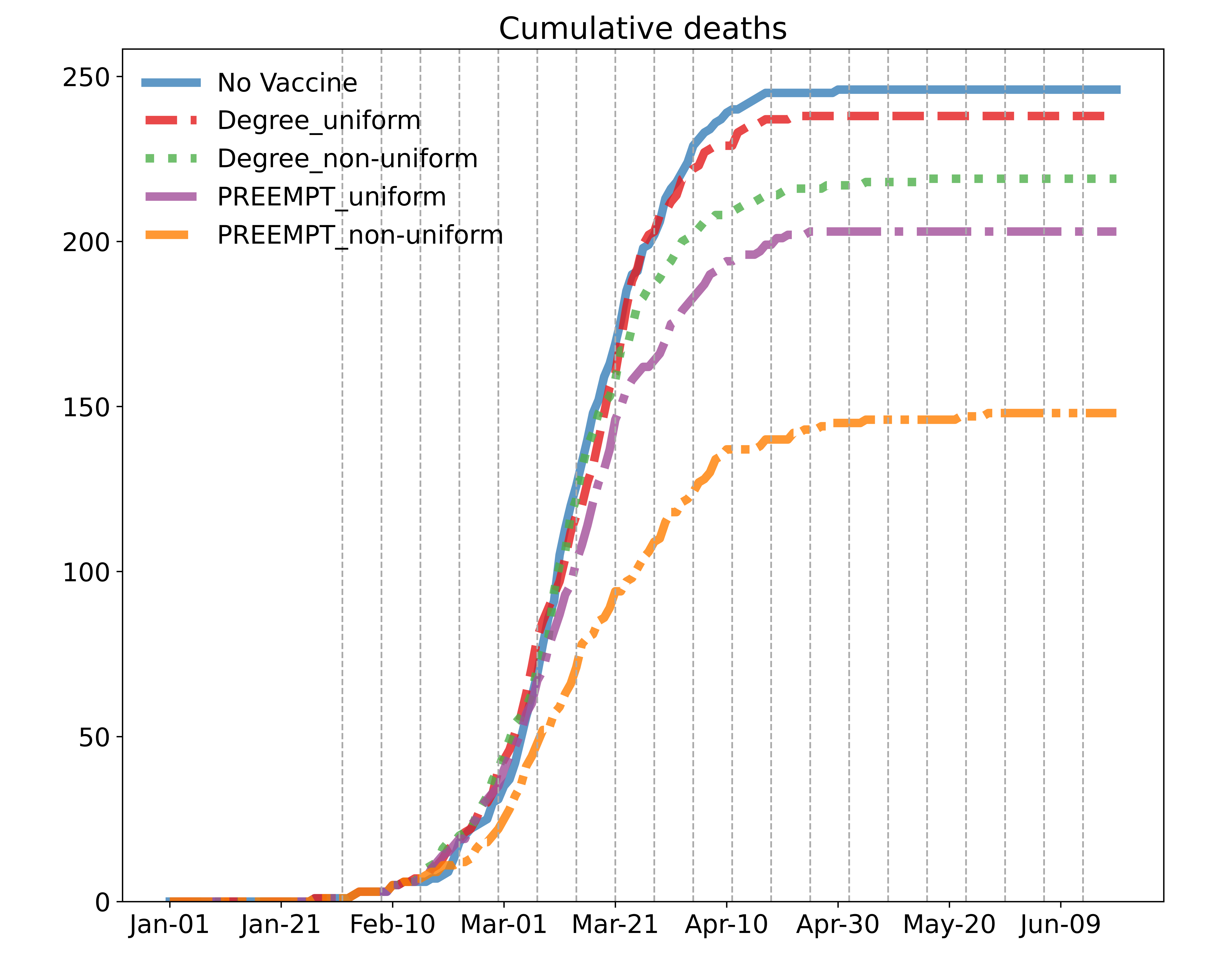}
         \caption{}
         \label{fig:5CD}
     \end{subfigure}
        \caption{
    \small
    \textit{Effects of varying the batch sizes as the disease spreads:} 
     The plots are labeled as `X\_Y' where `X' stands for the seed selection strategy and `Y' represents the two batching strategies---\textit{uniform} or \textit{non-uniform}. 
     The uniform strategy applies the same number of vaccines per round (1,000 vaccines per round). 
     The non-uniform strategy uses varying batch sizes. For these experiments, we used:
    2,000 vaccines in each of the first 5 rounds;  1,000 vaccines in each of the next 5 rounds; and 500 vaccines per round in the final 10 rounds.
     Also shown is the `No vaccine' curve for reference.}
        \label{fig:nonuniformvaccination}
        \vspace{-0.2cm}
\end{figure*}

We conducted a set of experiments setting a vaccination target of 20\% of the population over the simulation duration of $\sim$5 months. This is meant to reflect roughly the general pace at which various vaccination schemes are progressing in the \covid{} pandemic.
Over the 100,000 population, this translates to 20,000 vaccines in total. These vaccines are distributed across multiple rounds, and we use a \textit{uniform} batch size for these rounds. 

Figure~\ref{fig:uniformvaccination} shows the results of our experiments. Part (a) tracks the evolution of the number of cumulative infections; part (b) tracks the number of new infections per day; and part (c) tracks the number of cumulative deaths. 
We compare both the Degree and \preempt{} strategies, for two settings---i) using a single round with a batch size of 20,000; and ii) using 20 rounds with a batch size of 1,000. 
The key observations from Figure~\ref{fig:uniformvaccination} are as follows. 
First, we observe that the single round strategy (``20000\_1'') significantly outperforms the multiple round strategy (``1000\_20'').
This is to be expected; however, it is also not realistic to assume high vaccine capacity at the start of an epidemic. 
Second, we observe that \preempt{} consistently outperforms the corresponding Degree strategy under both settings (single and multiple rounds). 
In fact \preempt's performance with the multiple round setting (``1000\_20'') is comparable to the single round setting of Degree (``20000\_1''), in terms of both new infections per day and the number of cumulative deaths. 
These results collectively demonstrate the value of choosing seeds based on submodular optimization.

\subsubsection{Effect of vaccinating in batches of non-uniform sizes}
\label{sec:nonuniformbatches}

Next, we conducted a similar set of experiments where the batch sizes across vaccination rounds are non-uniform. There are two ways in which this strategy can be formulated---either as a ``top-heavy'' or a ``bottom-heavy''---based on which end of the time spectrum we use a larger batch size (top implies earlier batches, and bottom implies later batches). 
In both cases, we keep the total percentage of the population getting vaccinated the same. 
In this paper, we report results from the top-heavy strategy that reflects the real-world effort of aggressive vaccination at earlier stages of the pandemic.  
For this experiment, we used 2,000 vaccines in each of the first 5 rounds; 1,000 vaccines in each of the next 5 rounds; and 500 vaccines per round in the final 10 rounds.

The results can be seen in Figure~\ref{fig:nonuniformvaccination}. Like before, part (a) tracks the evolution of the number of cumulative infections; part (b) tracks the number of new infections per day; and part (c) tracks the number of cumulative deaths.
The key observations from Figure~\ref{fig:nonuniformvaccination} are as follows. The trends are similar across the three metrics, namely the number of cumulative infections, the number of new infections per day, and the number of cumulative deaths. 
Next, we observe that irrespective of batch size, \preempt{} remains the better strategy when it comes to seed selection in a batched setting, than the Degree variant. 
Finally, we also notice that when \preempt{} is used in non-uniform batch top heavy mode (``PREEMPT\_non-uniform''), it becomes significantly better in reducing the number of infections and deaths compared to even \preempt{} in uniform mode (``PREEMPT\_uniform''). 
This demonstrates the value of carefully choosing an appropriate subset of seeds early on in an epidemic.


\section{Conclusion} 

In this work, we provide a framework that integrates epidemic simulation with graph-theoretic/network science-based interventions.
Using this framework, users can select different subsets of seeds (nodes on the network) to vaccinate and observe their efficacies over the duration of the simulation time. Even though our paper uses a specific epidemic simulator to test, our framework itself is generic and can be integrated with other similar simulators. 

The experiments conducted in this paper help demonstrate a set of key findings. First, it shows the value of using an influence maximization based approach toward seed selection in the context of epidemic control. In particular, we demonstrate  that \preempt{} as a seed selection strategy is able to outperform other heuristics like degree or random schemes. 
In addition to a \textit{carefully selected} subset of seeds, our experiments also demonstrate that the \textit{timing} of these vaccination matter---i.e., giving more vaccines early on could save more lives in the long run. 
While a much broader set of experiments are necessary to establish the generality of these observations, these preliminary results point in a promising direction toward the value  of network-driven models for epidemic studies.

\section*{Acknowledgements}

The research is in parts supported by the U.S. DOE ExaGraph project at the Pacific Northwest National Laboratory (PNNL), and by the U.S. National Science Foundation (NSF) grants CCF 1815467, OAC 1910213, and CCF 1919122 to Washington State University.
PNNL is operated by Battelle Memorial Institute under Contract DE-AC06-76RL01830.
Any opinions, findings, and conclusions or recommendations expressed in this material are those of the author(s) and do not necessarily reflect the views of the funding agencies.

\bibliographystyle{plainnat}
\bibliography{references}

\begin{thebibliography}{17}
\providecommand{\natexlab}[1]{#1}
\providecommand{\url}[1]{\texttt{#1}}
\expandafter\ifx\csname urlstyle\endcsname\relax
  \providecommand{\doi}[1]{doi: #1}\else
  \providecommand{\doi}{doi: \begingroup \urlstyle{rm}\Url}\fi

\bibitem[Bearman et~al.(2004)Bearman, Moody, and Stovel]{bearman2004chains}
Peter~S Bearman, James Moody, and Katherine Stovel.
\newblock Chains of affection: The structure of adolescent romantic and sexual networks.
\newblock \emph{American journal of sociology}, 110\penalty0 (1):\penalty0 44--91, 2004.

\bibitem[Chen et~al.(2020)Chen, Levin, Eubank, Mortveit, Venkatramanan, Vullikanti, and Marathe]{chen2020networked}
Jiangzhou Chen, S~Levin, S~Eubank, H~Mortveit, S~Venkatramanan, A~Vullikanti, and M~Marathe.
\newblock Networked epidemiology for covid-19.
\newblock \emph{Siam news}, 53\penalty0 (5), 2020.

\bibitem[Cheng et~al.(2020)Cheng, Kuo, and Zhou]{cheng2020outbreak}
Chun-Hung Cheng, Yong-Hong Kuo, and Ziye Zhou.
\newblock Outbreak minimization vs influence maximization: an optimization framework.
\newblock \emph{BMC Medical Informatics and Decision Making}, 20\penalty0 (1):\penalty0 1--13, 2020.

\bibitem[Chhatwal et~al.(2021)Chhatwal, Ayer, and Linas]{covid19-sim}
Jagpreet Chhatwal, Turgay Ayer, and Benjamin~P. Linas.
\newblock {COVID-19 Simulator}, 2021.
\newblock URL \url{https://covid19sim.org/}.

\bibitem[Ferguson et~al.(2020)Ferguson, Nedjati~Gilani, and Laydon]{ferguson2020covid}
N~Ferguson, G~Nedjati~Gilani, and D~Laydon.
\newblock Covid-19 covidsim microsimulation model.
\newblock 2020.

\bibitem[Frias-Martinez et~al.(2011)Frias-Martinez, Williamson, and Frias-Martinez]{frias2011agent}
Enrique Frias-Martinez, Graham Williamson, and Vanessa Frias-Martinez.
\newblock An agent-based model of epidemic spread using human mobility and social network information.
\newblock In \emph{2011 IEEE third international conference on privacy, security, risk and trust and 2011 IEEE third international conference on social computing}, pages 57--64. IEEE, 2011.

\bibitem[Grefenstette et~al.(2013)Grefenstette, Brown, Rosenfeld, DePasse, Stone, Cooley, Wheaton, Fyshe, Galloway, Sriram, et~al.]{grefenstette2013fred}
John~J Grefenstette, Shawn~T Brown, Roni Rosenfeld, Jay DePasse, Nathan~TB Stone, Phillip~C Cooley, William~D Wheaton, Alona Fyshe, David~D Galloway, Anuroop Sriram, et~al.
\newblock Fred (a framework for reconstructing epidemic dynamics): an open-source software system for modeling infectious diseases and control strategies using census-based populations.
\newblock \emph{BMC public health}, 13\penalty0 (1):\penalty0 1--14, 2013.

\bibitem[Hethcote(2000)]{hethcote2000mathematics}
Herbert~W Hethcote.
\newblock The mathematics of infectious diseases.
\newblock \emph{SIAM review}, 42\penalty0 (4):\penalty0 599--653, 2000.

\bibitem[Kempe et~al.(2003)Kempe, Kleinberg, and Tardos]{DBLP:conf/kdd/KempeKT03}
David Kempe, Jon~M. Kleinberg, and {\'{E}}va Tardos.
\newblock Maximizing the spread of influence through a social network.
\newblock In Lise Getoor, Ted~E. Senator, Pedro~M. Domingos, and Christos Faloutsos, editors, \emph{Proceedings of the Ninth {ACM} {SIGKDD} International Conference on Knowledge Discovery and Data Mining, Washington, DC, USA, August 24 - 27, 2003}, pages 137--146. {ACM}, 2003.
\newblock \doi{10.1145/956750.956769}.
\newblock URL \url{https://doi.org/10.1145/956750.956769}.

\bibitem[Kermack and McKendrick(1927)]{kermack1927contribution}
William~Ogilvy Kermack and Anderson~G McKendrick.
\newblock A contribution to the mathematical theory of epidemics.
\newblock \emph{Proceedings of the royal society of london. Series A, Containing papers of a mathematical and physical character}, 115\penalty0 (772):\penalty0 700--721, 1927.

\bibitem[Kerr et~al.(2020)Kerr, Stuart, Mistry, Abeysuriya, Hart, Rosenfeld, Selvaraj, Nunez, Hagedorn, George, et~al.]{kerr2020covasim}
Cliff~C Kerr, Robyn~M Stuart, Dina Mistry, Romesh~G Abeysuriya, Gregory Hart, Katherine Rosenfeld, Prashanth Selvaraj, Rafael~C Nunez, Brittany Hagedorn, Lauren George, et~al.
\newblock Covasim: an agent-based model of covid-19 dynamics and interventions.
\newblock \emph{medRxiv}, 2020.

\bibitem[Kunst et~al.(2020)Kunst, Peralta, Reitsma, Andrews, Chin, Claypool, Covarrubias, Daniels, Fernandez, Fung, et~al.]{kunst2020stanford}
Andrea~Luviano Kunst, Yadira Peralta, Marissa Reitsma, Jose Manuel Cardona~Arias Andrews, Liz Chin, Anneke Claypool, Hugo~Berumen Covarrubias, Ally Daniels, Mariana Fernandez, Hannah Fung, et~al.
\newblock Stanford-cide coronavirus simulation model (sc-cosmo)--technical description document, version 2.0.
\newblock 2020.

\bibitem[Kuzdeuov et~al.(2020)Kuzdeuov, Baimukashev, Karabay, Ibragimov, Mirzakhmetov, Nurpeiissov, Lewis, and Varol]{kuzdeuov2020network}
Askat Kuzdeuov, Daulet Baimukashev, Aknur Karabay, Bauyrzhan Ibragimov, Almas Mirzakhmetov, Mukhamet Nurpeiissov, Michael Lewis, and Huseyin~Atakan Varol.
\newblock A network-based stochastic epidemic simulator: Controlling covid-19 with region-specific policies.
\newblock \emph{IEEE Journal of Biomedical and Health Informatics}, 24\penalty0 (10):\penalty0 2743--2754, 2020.

\bibitem[Minutoli et~al.(2020)Minutoli, Sambaturu, Halappanavar, Tumeo, Kalyanaraman, and Vullikanti]{DBLP:conf/sc/MinutoliSHTKV20}
Marco Minutoli, Prathyush Sambaturu, Mahantesh Halappanavar, Antonino Tumeo, Ananth Kalyanaraman, and Anil Vullikanti.
\newblock Preempt: scalable epidemic interventions using submodular optimization on multi-gpu systems.
\newblock In \emph{Proceedings of the International Conference for High Performance Computing, Networking, Storage and Analysis, {SC} 2020, Virtual Event / Atlanta, Georgia, USA, November 9-19, 2020}, page~55. {IEEE/ACM}, 2020.
\newblock \doi{10.1109/SC41405.2020.00059}.
\newblock URL \url{https://doi.org/10.1109/SC41405.2020.00059}.

\bibitem[Prem et~al.(2017)Prem, Cook, and Jit]{prem2017projecting}
Kiesha Prem, Alex~R Cook, and Mark Jit.
\newblock Projecting social contact matrices in 152 countries using contact surveys and demographic data.
\newblock \emph{PLoS computational biology}, 13\penalty0 (9):\penalty0 e1005697, 2017.

\bibitem[Schneider et~al.(2020)Schneider, Ngwa, Schwehm, Eichner, and Eichner]{schneider2020covid}
Kristan~A Schneider, Gideon~A Ngwa, Markus Schwehm, Linda Eichner, and Martin Eichner.
\newblock The covid-19 pandemic preparedness simulation tool: Covidsim.
\newblock \emph{BMC infectious diseases}, 20\penalty0 (1):\penalty0 1--11, 2020.

\bibitem[Tang et~al.(2015)Tang, Shi, and Xiao]{DBLP:conf/sigmod/TangSX15}
Youze Tang, Yanchen Shi, and Xiaokui Xiao.
\newblock Influence maximization in near-linear time: {A} martingale approach.
\newblock In Timos~K. Sellis, Susan~B. Davidson, and Zachary~G. Ives, editors, \emph{Proceedings of the 2015 {ACM} {SIGMOD} International Conference on Management of Data, Melbourne, Victoria, Australia, May 31 - June 4, 2015}, pages 1539--1554. {ACM}, 2015.
\newblock \doi{10.1145/2723372.2723734}.
\newblock URL \url{https://doi.org/10.1145/2723372.2723734}.

\end{thebibliography}

\end{document}